\documentclass[%
 aip,
 amsmath,amssymb,
 reprint,%
]{revtex4-1}

\usepackage{graphicx}
\usepackage{dcolumn}
\usepackage{bm}

\usepackage{hyperref}

\usepackage{color}

\usepackage[utf8]{inputenc}
\usepackage[T1]{fontenc}
\usepackage{mathptmx}
\usepackage{upgreek}

\begin{document}


\title{A Particle-In-Cell Code Comparison for Ion Acceleration: EPOCH, LSP, and WarpX}

\author{Joseph R. Smith} \email{smith.10838@osu.edu}
 \affiliation{Department of Materials Science and Engineering, The Ohio State University, Columbus, Ohio 43210, USA}

\author{Chris Orban}%
\affiliation{ 
Department of Physics, The Ohio State University, Columbus, Ohio 43210, USA
}%

\author{Nashad Rahman}%
\affiliation{ 
Department of Physics, The Ohio State University, Columbus, Ohio 43210, USA
}%

\author{Brendan McHugh}%
\affiliation{ 
Department of Physics, The Ohio State University, Columbus, Ohio 43210, USA
}%

\author{Ricky Oropeza}%
\affiliation{ 
Department of Physics, The Ohio State University, Columbus, Ohio 43210, USA
}%

\author{Enam A. Chowdhury}
 \affiliation{Department of Materials Science and Engineering, The Ohio State University, Columbus, Ohio 43210, USA}
  \affiliation{ 
Department of Electrical and Computer Engineering, The Ohio State University, Columbus, Ohio 43210, USA
}%
 \affiliation{ 
Department of Physics, The Ohio State University, Columbus, Ohio 43210, USA
}%

\date{\today}

\begin{abstract}
There are now more Particle-in-Cell (PIC) codes than ever before that researchers use to simulate intense laser-plasma interactions. To date, there have been relatively few direct comparisons of these codes in the literature, especially for relativistic intensity lasers interacting with thin overdense targets. To address this we perform a code comparison of three PIC codes:  EPOCH, LSP, and WarpX for the problem of laser-driven ion acceleration in a 2D(3v) geometry for a $10^{20}$ W~cm$^{-2}$ intensity laser. We examine the plasma density, ion energy spectra, and laser-plasma coupling of the three codes and find strong agreement. We also run the same simulation 20 times with different random seeds to explore statistical fluctuations of the outputs. We then compare the execution times and memory usage of the codes (without ``tuning'' to improve performance) using between 1 and 48 processors on one node.  We provide input files to encourage larger and more frequent code comparisons in this field.

\end{abstract}

\maketitle

The Particle-in-Cell (PIC) method has origins tracing back to early work by Dawson published in 1962\cite{dawson1962one,hockney1981PIC,BirdsallLangdon2004}, and has since become a ubiquitous tool to model intense laser-plasma interactions (LPI). PIC is used both to predict the outcome of experiments and in \emph{post facto} modeling of experiments that were performed in order to understand LPI at spatial and temporal scales that are difficult or impossible to achieve with standard diagnostics.
Code comparisons play an important role in fields where there are relatively few analytically (or semi-analytically) solvable problems to compare codes against. In hydrodynamics, for example, there have been a number of studies comparing codes for various problems of interest to astrophysics (e.g., \cite{Frenk_etal1999,Tasker_etal2008,Robertson_etal2010,Vazza_etal2011}) and a handful of studies in high energy density physics\cite{Fatenejad_etal2013,Orban_etal2020}.
Code comparisons are also regularly performed for codes that simulate non local thermal equilibrium (NLTE) radiation physics going back to 1996 \cite{nlte}.

There is much more limited work comparing PIC codes for LPI, which is unfortunate since many PIC codes have been developed or are in development to model LPI. An incomplete list is AlaDyn~\cite{ALaDyn_benedetti2008tt}, Chicago~\cite{Chicago_Thoma_2017}, EPOCH~\cite{arber2015contemporary}, FBPIC~\cite{LEHE201666}, LSP~\cite{Welch_etal2004,Welch_etal2006}, OSIRIS~\cite{OSIRIS}, PICCANTE~\cite{piccante_sgattoni2015optimising}, PICLS~\cite{SENTOKU2008_PICLS}, PIConGPU~\cite{PIConGPU2013}, PSC~\cite{PSC_germaschewski2016plasma}, Smilei~\cite{derouillat2018smilei},  VLPL~\cite{pukhov1999threelvpl}, VORPAL~\cite{VORPAL_nieter2004vorpal}, VPIC~\cite{bowers20080_VPIC}, and WarpX~\cite{VAY2018476}. These codes are written in a variety of programming languages, include different physics packages and numeric algorithms, and have various degrees of availability  (e.g., commercial, requiring a formal/informal agreement, or open source).

Prior LPI code comparisons have primarily focused on laser wakefield acceleration (LWFA)\cite{paul2009benchmarking,ALaDyn_benedetti2008tt, 
VAY2018476} and are limited in scope. There have also been efforts comparing algorithms for a given PIC code~\cite{Cowan_PhysRevSTAB.16.041303,VAY2014610} and other relevant benchmarking and code-comparison work outside of LPI~\cite{woods2010magic,turner_2013_doi:10.1063/1.4775084,turner2016verification,Riva_doi:10.1063/1.4977917,oconnnor_3d}. For ion acceleration, \citet{cochran2018new} compares the explicit and implicit algorithms in LSP in a PhD dissertation. \citet{Tazes_2020} compare EPOCH and PIConGPU for ion acceleration, although the focus is primarily on performance for CPU and GPU architectures. \citet{mouziouras_compare_thesis} performs a direct comparison of EPOCH and Smilei for ion acceleration in a Master's thesis. In our work, we compare three codes: EPOCH (university code, open source), LSP (commercial code), and WarpX (laboratory code, open source), with a straightforward 2D(3v) test problem of ion acceleration. We compare both simulation results and the performance of the codes.


The simulation setup for these 2D(3v) simulations is summarized in Fig.~\ref{fig:sketch} and Tab.~\ref{tab:param}. The EPOCH and WarpX input `decks' are provided in the supplementary material to encourage similar comparison in the future.  The $\lambda$ = 800~nm wavelength laser propagates in the $+x$ direction with linear polarization out of the simulation plane\cite{stark2017effects}\footnote{The EPOCH simulations have an $xy$ geometry, which is converted to $xz$ for consistency.}. The laser has a Gaussian spatial intensity profile with a 3 $\upmu$m Full-Width-at-Half-Maximum (FWHM) spot size (beam waist $w_0\approx2.55~\upmu$m), peak intensity of 10$^{20}$~W~cm$^{-2}$, and a sine-squared temporal profile with a 30~fs FWHM pulse duration. For this intensity laser, the normalized vector potential $a_0 = {eE_0 }/{ m_e \omega c}$ = 6.8, is in the relativistic regime ($a_0 \gtrsim 1$), where $e$ is the electron charge, $E_0$ is the peak electric field amplitude, $m_e$ is the electron mass, $\omega = 2\pi c/\lambda$ is the laser angular frequency, and $c$ is the speed of light. The laser arrives at normal incidence to the target which is composed of ionized hydrogen at a density of $8.5\times 10^{21}$ cm$^{-3} \approx 5n_{crit}$, where $n_{crit} = {\varepsilon_0 m_e \omega^2}/{e^2}$, is the nonrelativistic critical density of the plasma, where $\varepsilon_0$ is the permittivity of free space. This relatively low density is chosen for computational convenience and is barely above critical density, particularly if relativistic effects are considered~\cite{Wilks_etal1992,Haines_2009_PhysRevLett.102.045008}. The target is 1~$\upmu$m thick in  $x$  and 20~$\upmu$m in  $z$. These parameters allow us to explore the well-studied Target Normal Sheath Acceleration (TNSA) mechanism~\cite{hatchett2000tnsa,snavely_2000_tnsa,Mora2003,Fuchs_etal2009,macchi_ion_2013,Wagner2016TNSA85MeV} in a regime where collisional effects are expected to be minimal\cite{Kemp_etal2004}.

\begin{figure}
\includegraphics{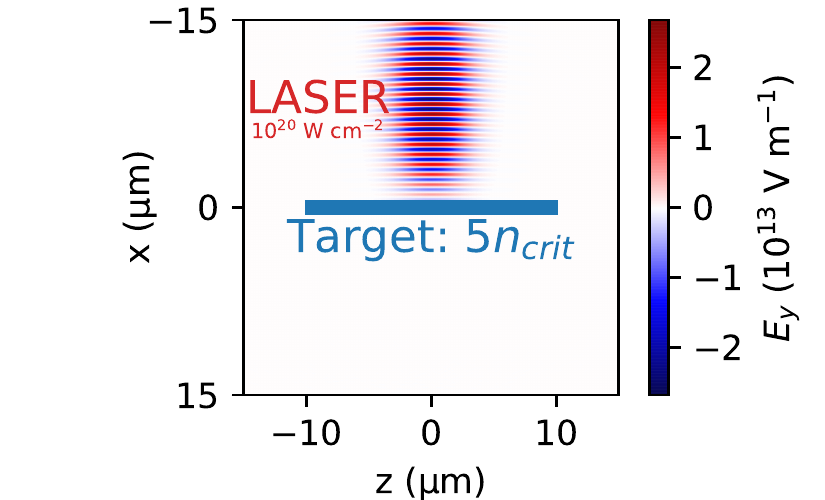}
\vspace{-0.3cm}
\caption{The simple 2D simulation setup, where a linearly polarized laser (in the virtual $y$ dimension),  interacts with a 1~$\upmu$m thick target.  } \label{fig:sketch}
\end{figure}

\begin{table}
\caption{Summary of simulation parameters.}\label{tab:param}
\begin{ruledtabular}
\begin{tabular}{l|l|l}
Laser&Wavelength; Peak Intensity & 800~nm; $10^{20}$ W~cm$^{-2}$ \\
&Pulse Duration; Spot Size  & 30 fs FWHM; 3 $\upmu$m~FWHM \\

\hline
Target&Material & Ionized Hydrogen \\
& Density &$8.5\times 10^{21}$ cm$^{-3}$ \\

\hline
Sim.&Resolution & 20~nm \\
&Timestep Size; Total Time & 0.04~fs; 300~fs  \\
& Macroparticles Per Cell & 100 Electrons, 100 Protons \\
& Initial Particle Temperature & 10~keV  \\
\end{tabular}
\end{ruledtabular}
\end{table}

The grid is divided into 20~nm cells with 100 macroparticles per cell per species (ten million total particles, 1500 $\times$ 1500 cells). The particles are initialized with a temperature of 10~keV. The initial Debye length $\lambda_D = \sqrt{\epsilon_0 k_B T_e/n_ee^2}$ = 8~nm, where $k_B$ is the Boltzmann constant, $T_e$ is the electron temperature, and $n_e$ is the electron density. For explicit codes the Debye length must be sufficiently resolved to limit numerical heating~\cite{Numerical_Heating_Horky_1027}, where self-heating rates depend on a code's shape function and number of macroparticles per cell~\cite{arber2015contemporary}. We also consider the plasma frequency $\omega_{p} =\sqrt{n_e e^2/m_e\varepsilon_0}$, where the skin depth~\cite{macchi2013superintense} is $l_s \approx c/\omega_p$ = 58~nm. We use $\approx$ 3 cells per skin depth, which is important to resolve for proper absorption~\cite{Kemp_skin_depth}. 


The timestep is a constant 0.04~fs ($\approx0.848$ times the CFL limit) and the simulation is run for 300~fs (7500 steps). Explicit algorithms require the CFL number to be less than one for numerical stability, whereas implicit approaches can potentially exceed this limitation. We stop the simulation at 300~fs; at this time the total ion energy has begun to plateau, yet no ions have left the simulation grid. Given a sufficiently large simulation box, the maximum ion energy can continue to slightly increase for hundreds of fs. Simple outlet/outflow/Silver-M\"uller boundary conditions are employed. The laser is introduced at the $x=-15~\upmu$m boundary for EPOCH and LSP. The focal spot location is at $x=0~\upmu$m, $z=0~\upmu$m.  WarpX uses an antenna model for the laser\footnote{See the documentation~\protect\url{https://warpx.readthedocs.io/} for more information.}. We place the antenna inside the grid, 10~nm from the boundary  (at $x=-14.99~\upmu$m, $z=0$ ~$\upmu$m) to allow easy comparison of the outputs and limit potential interactions with the boundary.

In this comparison, we use LSP version 10, EPOCH 4.17.10, and WarpX commit \texttt{a6fdf15}\footnote{\url{https://github.com/ECP-WarpX/WarpX/commit/a6fdf159cc15c46e5c2fbe711086b4e094935d55}} (between tagged versions 21.03 and 21.04). Simulations were designed to be as similar as possible to each other while generally following the default settings and limitations of each code. Both EPOCH and WarpX are explicit codes, whereas LSP has both explicit and implicit algorithms. We consider both approaches for LSP and refer to them as LSP(E) and LSP(I) respectively. LSP can operate as a hybrid PIC code with fluid physics, although those options are not evaluated here. We do not consider collisions, ionization, or any other effects, although these would be interesting to explore in future comparisons. The physical and numeric choices for the problem are not set to represent any actual experiment, but rather to provide a relatively simple benchmark to compare the physics in these codes that can feasibly be tested on a single node of a cluster. To limit the computational requirements we use 2D(3v) simulations instead of more realistic 3D simulations. We note that there are well-documented differences between the predictions of 2D(3v) and 3D PIC simulations for ion acceleration \cite{LIU_2013_2D_3D,Ngirmang_etal2016,stark2017effects,2D_3D_Babaei_2017,Xiao_2018_2D3D}.


Modern PIC codes commonly suggest higher order particle shape functions for ion acceleration simulations. By default, EPOCH employs a `second-order' triangle shape function, where a macroparticle has a width of two times the grid spacing and overlaps 3 cells in each dimension (see \citet{SHARP_Shalaby_2017} for discussion of the common `shape' and corresponding `weight' functions found in PIC codes). We also use this second-order shape function for WarpX, although for LSP we use a `first-order' cloud-in-cell algorithm; this is the highest order we found in this version of the code. We note that some authors\cite{arber2015contemporary} follow a different convention for the `order' of a shape function than the one used here~\cite{BirdsallLangdon2004,SHARP_Shalaby_2017}. By default, for these configurations, LSP and WarpX use an energy-conserving field gathering approach whereas EPOCH uses a more classical momentum-conserving approach\footnote{EPOCH has an option for current smoothing, but this is not used by default nor considered here}.

\begin{figure}
\includegraphics{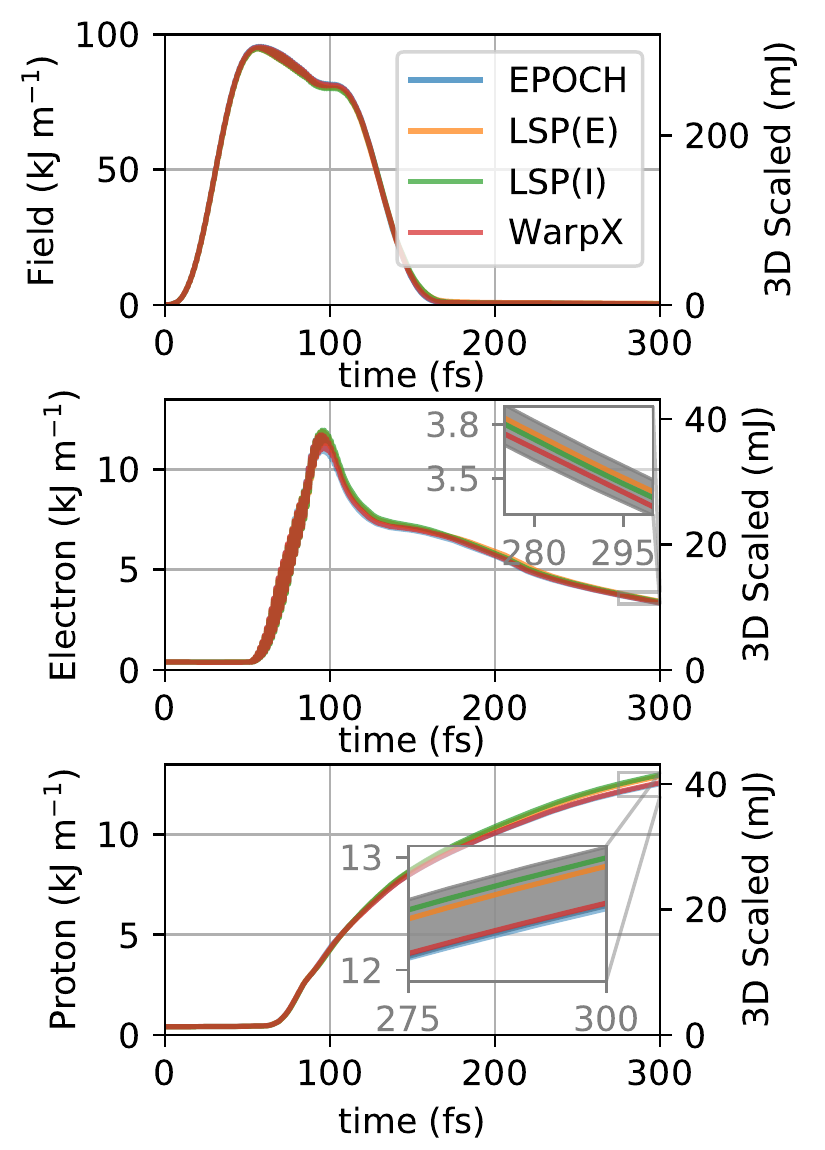}
\vspace{-0.3cm}
\caption{The total field (top), electron (middle), and proton (bottom) energies throughout the simulations. The left axis reports the energy in kJ assuming a 1~m virtual dimension, the right axis  in mJ is multiplied by $w_0 \sqrt{\pi/2}$ to facilitate comparisons to 3D simulations and experiments. The shaded regions in the inset boxes represent the range of energies (sampled every 10 fs) from the set of 20 EPOCH simulations with different random seeds.  [Associated dataset available at \url{https://doi.org/10.5281/zenodo.4651296}. ] (Ref.~\cite{dataset_zenodo}). } \label{fig:energyPlots}
\end{figure}

We compare the field, electron, and proton energy on the simulation grid in Fig.~\ref{fig:energyPlots}. Due to the 2D(3v) geometry, energy is calculated considering a 1~meter `virtual' $y$ dimension in EPOCH and WarpX, whereas LSP uses 1~cm by default.  We have also scaled the energies according to the spot size of the laser for comparison to 3D simulations and experiments with these laser parameters.  Figure~\ref{fig:energyPlots} shows good agreement between the energy diagnostics in all of the codes at early times ($ \lesssim 50$~fs) when the laser is propagating on the grid without interacting with any particles. The simulations remain highly consistent as the laser interacts with the target and transfers energy to the particles. At the end of the simulation the total electron and ion energies of the various simulations agree within a relative difference of 4\%. This is within reasonable fluctuations, as demonstrated by running 20 additional EPOCH simulations with different random seeds as shown in the inset boxes in Fig.~\ref{fig:energyPlots}. LSP provides an estimate for energy conservation in the simulation, with LSP(E) gaining less than 0.1\% and LSP(I) losing less than 0.2\% of the total energy.


\begin{figure}
\includegraphics{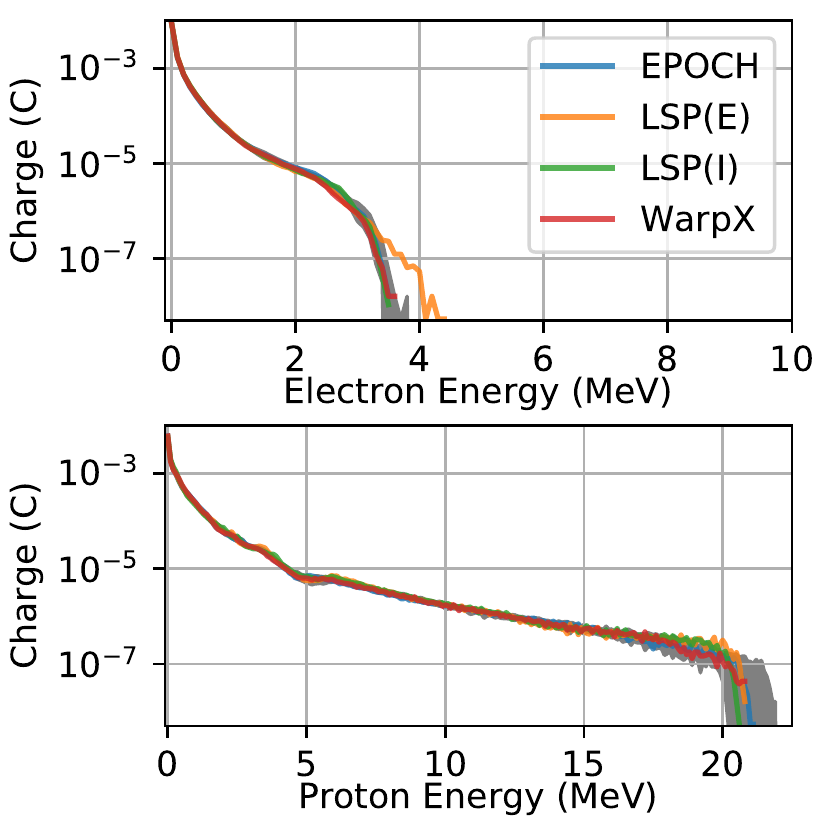}
\vspace{-0.5cm}
\caption{Distribution of forward going (velocity in $+x$) electrons (top) and ions (bottom) at 300~fs with a bin size of 0.1~MeV. The supplemental video (Multimedia view) shows the spectra at many timesteps. The shaded region represents the spread from 20 EPOCH simulations with different random seeds.  [Associated dataset available at \url{https://doi.org/10.5281/zenodo.4651296}. ] (Ref.~\cite{dataset_zenodo}). } \label{fig:2DdSpec}
\end{figure}

Figure~\ref{fig:2DdSpec} shows the electron and ion energy spectrum for each simulation, which are in good agreement. The maximum ion cut-off energy ranges from 20.6~MeV for LSP(I)  to  21.1~MeV for EPOCH. This measurement is being made from a relatively small number of macroparticles at these high energies so it is understandable that the codes disagree at this level as illustrated by the shaded region from many simulations with different seeds. This is also generally within expected experimental uncertainty.

There are subtle differences in the electron spectra early in the laser-plasma interaction ($\sim$60~fs) potentially due to differences in the laser generation and propagation for the codes. At late times ($\gtrsim$230~fs), the electron spectra for the LSP(E) simulation exhibits higher energies. These differences should be probed further in future work, although they seem to have a limited effect on the final ion energy distribution. We did find a difference in the number of electron macroparticles leaving the simulations, where the 21 EPOCH runs, and LSP(I)/WarpX simulations had $0.67-0.70$\% leave and LSP(E) had 0.82\% leave.


\begin{figure*}
\includegraphics{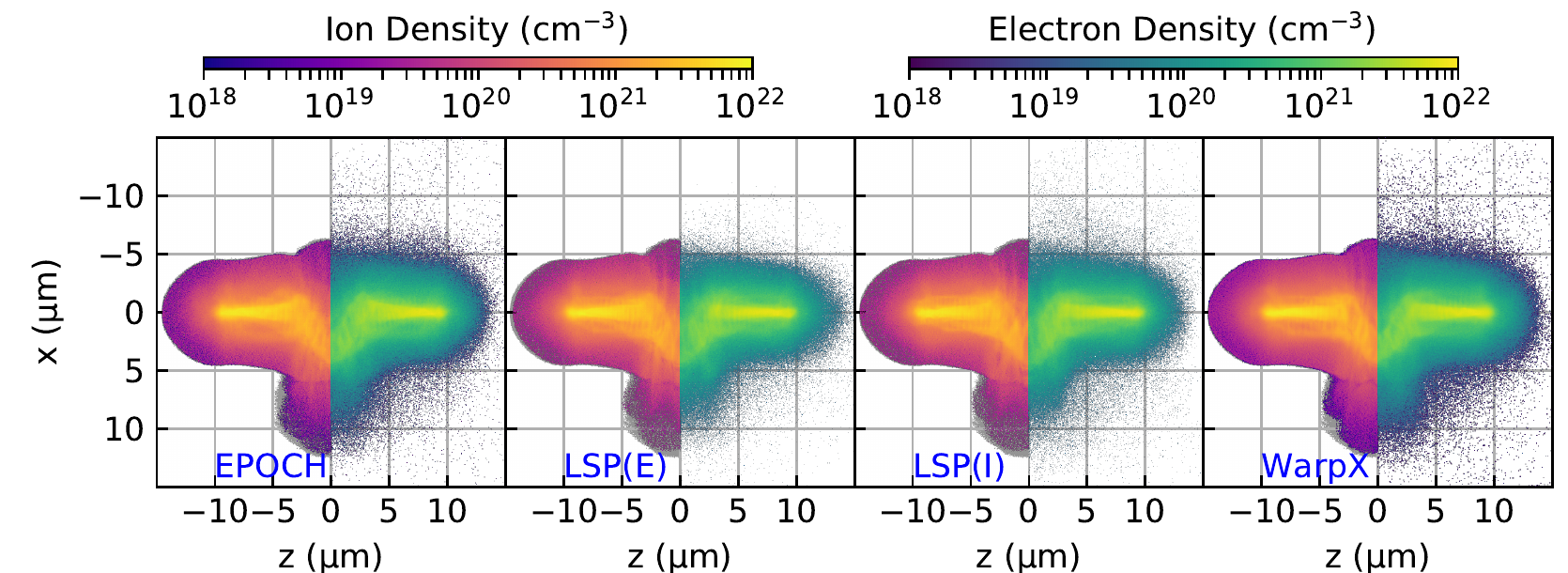}
\vspace{-0.3cm}
\caption{Ion ($z<0$) and electron ($z>0$) densities of the simulations at 300~fs. In the included video, the evolution of the interaction is included, where the laser is represented by a contour drawn when the magnitude of the electric field is 1/$e$ of the maximum value for this laser at focus in vacuum, or $E \approx 10^{13}$~V m$^{-1}$  (Multimedia view). The codes show good agreement, where the gray regions denote cells in which at least one of the additional simulations had a non-zero ion density. The increased granularity of the LSP plots is due to the lower order shape function.  } \label{fig:2Ddens}
\end{figure*}

\begin{figure}
\includegraphics{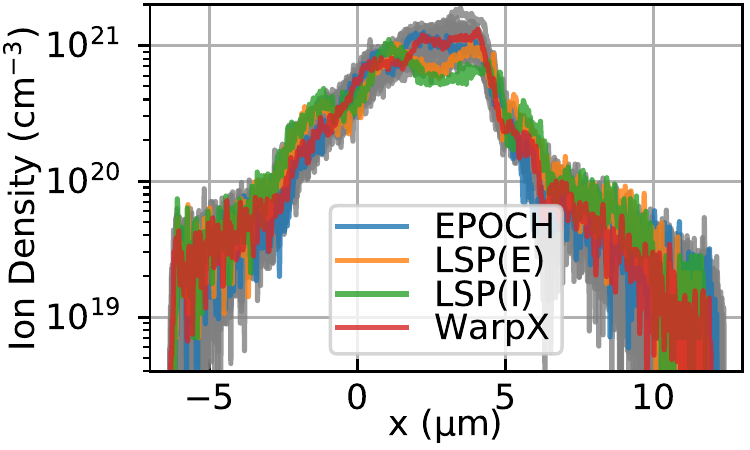}
\vspace{-0.3cm}
\caption{A lineout (averaged over 6 cells) of the final ion density through the center of the target for the codes and 20 additional runs. } \label{fig:lineout}
\end{figure}

In Fig.~\ref{fig:2Ddens} we present the final electron and ion density distributions. The final density distribution shapes are qualitatively similar, although the shape of the high energy tails are varied and there are distinct differences in the actual expanded target density, although we note that the color scale is logarithmic and the final profile is susceptible to the initial particle positions as shown in Fig.~\ref{fig:2Ddens}. There appear to be fewer electrons for LSP(E) $x\lesssim -10~\upmu$m, where more particles have left the grid in that direction. The lower order shape function in LSP results in a more granular density profile. To make a more quantitative comparison, Fig.~\ref{fig:lineout} shows a lineout of ion density at the end of the simulations.

\begin{figure}
\includegraphics{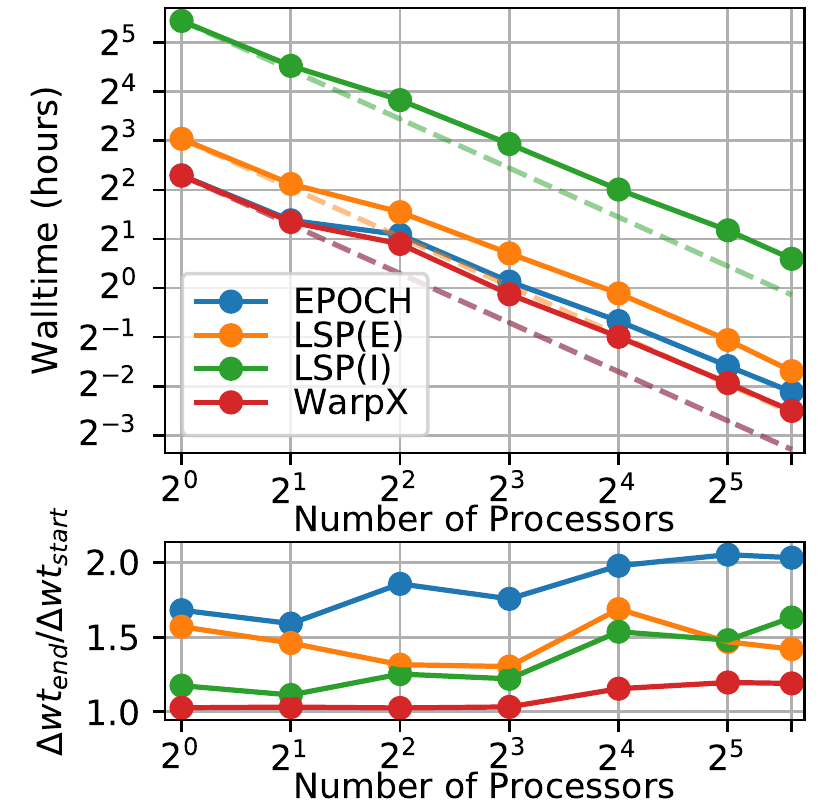}
\vspace{-0.2cm}
\caption{Scaling results for 1 to 48 processors on a single node (top). Ideal scaling is represented with dashed lines. A simple static grid division scheme is used, where the grid is divided into strips along~$z$. Load balancing for $\geq 4$ processors is non-ideal since the target has a finite  extent. On the bottom we divide the average wall-time duration of timesteps at the end of the simulation (last 100 steps) by the beginning (first $\sim 100$ steps). 
All of the codes run slower towards the end of the simulation with WarpX having the most uniform performance.
} \label{fig:scaling}
\end{figure}

We also compared the execution time and scaling of the codes at the Ohio Supercomputer Center's\cite{OhioSupercomputerCenter1987} Pitzer cluster on nodes with 48 cores (dual Intel Xeon 8268 processors), with Intel 19.0.5 compilers and MVAPICH2 2.3.3. We focused on the compute time of each code and did not write particle or field data to disk during timing runs. Each code internally tracked and logged timing statistics. First we look at single core performance, where the simulation grid is not divided as the simulations are run on a single core. This provides a comparison of the raw performance of each code without taking different parallelism strategies into account. As shown in Fig.~\ref{fig:scaling}, WarpX and EPOCH had the best single core performance this followed by LSP(E), which required nearly twice the computation time. The average wall-time per particle per push was $2.3\times 10^{-7}$~s for EPOCH and WarpX, $4.0\times 10^{-7}$~s for LSP(E), and $2.1\times 10^{-6}$~s for LSP(I)\footnote{These calculations are made assuming no particles leave the simulation, corrections would be minimal as < 0.5\% of the total particles leave by the end of each simulation.}. This difference is especially interesting since, as mentioned, the shape function for LSP was lower order than WarpX or EPOCH, which should decrease the  total computations. Using a higher order shape function would have increased the computation time for the LSP runs making this difference even larger. LSP(I) took about 9 times longer than EPOCH or WarpX, which we note is not entirely surprising for the implicit algorithm.   

When testing the codes with multiple processors on a single node using MPI for all codes\footnote{The same executable was used for single and multi-processor runs. WarpX has both MPI and OpenMP implemented, although for this problem/system OpenMP did not seem to improve performance, so this option was disabled during compilation for these tests. }, each simulation box is divided into equal (or nearly equal) strips along $z$ (parallel to the $x$ axis) with full extent in the $x$ direction. This was readily compatible with the various approaches in each code. As shown in Fig.~\ref{fig:scaling}, we see nearly ideal scaling to two processors, which unsurprising as the domain and particles can be evenly divided in two. With four processors this simple grid division scheme produces an uneven load balance due to the finite transverse extent of the target. This is not ideal, but still provides useful information comparing the performance of the codes. For additional processors, WarpX performed the best, followed by EPOCH and LSP(E) and LSP(I). At early times the EPOCH code ran more quickly per timestep than WarpX, but at late times the EPOCH timesteps ran less quickly than WarpX which had a relatively constant wall-time per timestep throughout the simulations as shown in Fig.~\ref{fig:scaling}.

We can make a simple lower-bound estimate for memory usage. Each cell requires 9 variables (3 each for $\vec E$, $\vec B$, and $\vec J$), which all need 8 bytes, assuming double precision\cite{hubl2019picongpu}. We have $1500 \times 1500$ cells $\times$ 72 bytes/cell = 146 MB. At minimum in a 2D(3v) geometry, particles need 2 variables for position, 3 for momentum, and 1 for particle weighting\cite{hubl2019picongpu,macchi2013superintense}\footnote{Some codes provide the option for a per-species particle weight, rather than per particle. We do not consider this option.}, or 48 bytes/particle. Memory requirements will vary based on the data structures in the code (e.g.~\citet{VPIC_2.0} utilize lower precision variables to save memory) and codes may use more variables per particles. We have 10,000,000 particles $\times$ 48 bytes/particle = 480 MB. This provides a lower-bound of $\approx$ 626 MB. For 1 core runs, the XDMoD metric tool~\cite{XDMod} reports maximum memory usage of EPOCH: 1.4~GB, LSP(E): 1.3~GB, LSP(I): 7.8~GB, and WarpX: 1.2~GB. 



We caution the reader from concluding that this is a definitive ranking for the performance of the codes. As already mentioned, we did not write particle or field data to disk during these performance tests which means that our results do not speak to the efficiency of the parallel I/O in any of the codes. Also, the execution time of the codes could be improved by more intelligent domain decomposition and/or dynamic load balancing. We did not try to optimize the execution time of the various codes but certainly this is worthy of attention in future work.

We find a high level of agreement between three popular codes for simulating ion acceleration. We found better than 4\% agreement in final particle energy in the simulations and similar cutoff energies for the ion energy spectrum. We also compared the results from the codes to 20 additional EPOCH simulations that were performed with different random seeds. This was to quantify differences that may arise because the of the finite number of macroparticles. Broadly we found that the differences between the codes, including the high energy tail of the ion energy distribution, fell within the range of these 20 additional simulations.  Identifying systematic differences in simulation results due to distinct algorithmic and numeric choices in the codes would benefit from higher resolution tests.

A comparison of code performance revealed, as expected, that the LSP code in implicit mode was slower and consumed more memory than the other explicit codes (and LSP in explicit mode). One expects an implicit code to consume more memory and run more slowly than an explicit code because the program holds on to the electric and magnetic fields both at the current step and a half step into the future. These results highlight that for some problems a much higher resolution explicit simulation may be more efficient than a lower resolution implicit simulation.


We do not claim that any one of the codes we considered is more accurate than another. The choice of code often depends strongly on the problem being solved and the computational resources available. For example, the implicit solver in LSP can be useful for modeling lower temperature plasmas with reduced numerical effects and can produce stable results with larger grid sizes and timesteps, reducing the computation time. LSP also includes many different physics packages (e.g.,~a hybrid fluid model and a circuit model) not yet found in the other codes. WarpX and LSP both offer some non-uniform grid capabilities, which can improve execution time, and WarpX has GPU support. 
Clearly, there are many distinct features of different codes, and new features and physics packages are being actively developed. We encourage the reader to consult the documentation, user guides, recent publications and the developers themselves for the most up-to-date features and capabilities of each code and to consider the many other excellent PIC codes not used here. There are many additional features of these codes that can be compared such as collisions and ionization that should be explored with future work.

LPI simulations typically show good agreement with experiments, although they often depend on assumptions about pre-pulse/pre-plasma effects that may not match the experiment. Future experiments used for validation of these codes can focus on careful matching of the initial conditions and accurate quantification of the plasma properties throughout the experiment. Validation tests need not be limited to LPI. For example, PIC codes can also be used to model experiments studying capacitively coupled plasmas (e.g.,~\cite{Rauf_2020,Wang_2021_cpc}), and conditions related to plasmas created during hypersonic flight/reentry~\cite{usui2000computer,Thoma_2007_hyper,Krishnamoorthy_2017}.


In conclusion, PIC simulations are an essential tool to model and understand laser-plasma interactions but thus far there has been limited work comparing codes to one another. We find strong agreement among EPOCH, LSP and WarpX codes with only minor differences in electron and ion densities and in the high energy cutoff that can largely be attributed to the finite number of macroparticles.
We also show important differences in performance among the codes, which of concern for efficient resource utilization. We hope that this work encourages LPI code comparisons in the future. 

\section*{Supplementary Material}
See supplementary material for the EPOCH and WarpX input files for these simulations. 

\begin{acknowledgments}
This research was funded by DOE STTR grant no.~DE-SC0019900. This project utilized resources from the Ohio Supercomputer Center~\cite{OhioSupercomputerCenter1987}. We would like to thank Axel Huebl, R\'emi Lehe, and Jean-Luc Vay from the WarpX development team for useful discussions. We would also like to thank Tony Arber and the EPOCH development team for helpful feedback. We also thank Gregory Ngirmang for useful discussions. The code EPOCH used in this work was in part funded by the UK EPSRC grants EP/G054950/1, EP/G056803/1, EP/G055165/1 and EP/ M022463/1. This research used the open-source PIC code WarpX \url{https://github.com/ECP-WarpX/WarpX}, primarily funded by the US DOE Exascale Computing Project. We acknowledge all WarpX contributors. Figures in this paper were generated using matplotlib~\cite{Hunter:2007}.

\end{acknowledgments}

\section*{DATA AVAILABILITY}
Much of the data that support the findings of this study are openly available in Zenodo at \url{https://doi.org/10.5281/zenodo.4651296}\cite{dataset_zenodo}. Other data that support the findings of this study are available from the corresponding author upon reasonable request.




\bibliography{main}

\end{document}